\documentclass{cimento}
\title{Feedback-assisted measurement of the free mass position 
\ETC \ under the Standard Quantum Limit}
\author{G. M. D'Ariano, M. F. Sacchi and R. Seno}
\instlist{
\inst Dipartimento di Fisica \lq Alessandro Volta\rq and INFM---Unit\`a di
PAVIA,\\Universit\`a degli Studi di Pavia \\
via A. Bassi 6, I-27100 Pavia, Italy}
\PACSes{\PACSit{03.65.Bz, 42.50.Dv, 42.50.-p}{}}

\begin{document}

\maketitle

\begin{abstract}
The Standard Quantum Limit (SQL) for the measurement of a free mass position 
is illustrated, along with two necessary conditions for breaching
it. A measurement scheme that overcomes the SQL is engineered.
It can be achieved in three-steps: {\em i)} a pre-squeezing stage; 
{\em ii)} a standard von Neumann measurement with momentum-position
object-probe interaction and {\em iii)} a feedback. Advantages and 
limitations of this scheme are discussed. It is shown that all of the three
steps are needed in order to overcome the SQL. In particular, the von
Neumann interaction is crucial in getting the right state reduction, 
whereas other experimentally achievable Hamiltonians, as, for example,
the radiation-pressure interaction, lead to state reductions that on the 
average cannot overcome the SQL.
\end{abstract}      

\section{Introduction}\label{s:introduction}

The problem of achieving a sequence of measurements of the position of
a free mass with arbitrary precision has received much attention in
the past, especially as a tool for monitoring the presence of an
external classical field weakly interacting with the mass
itself---typically, gravitational waves \cite{bragi74,edelst78}.
Originally, a {\em Standard Quantum Limit} (SQL) was devised
\cite{bragi74,caves80} stating that {\em if a mass evolves freely for
a time interval $t_f$ between two measurements, the uncertainty of the
second measurement cannot be lower than} $\Delta^2_{SQL}=\hbar t_f/m$,
$m$ being the inertial mass of the freely moving body.  \par The SQL
can be shortly illustrated as follows. Let us consider that the moving
mass after the first measurement at $t=0$ is described by the
following minimum uncertainty wave-packet (MUW) centered at $q_0$, and
moving to the right with momentum $\hbar k_0$
\begin{eqnarray}
\psi(q,0)=\left( \frac{1}{2\pi\delta^2}\right )^{\frac{1}{4}}
\exp\left[-\frac{(q-q_0)^2}{4\delta^2 }+ik_0 (q-q_0)\right]
\;.\label{e:muwinit}
\end{eqnarray}  
If such wave-packet undergoes a free evolution for a time $t_f$, its 
initial position variance $\delta^2=\langle \Delta\hat q^2(0)\rangle $ 
increases as follows \cite{schiff}
\begin{eqnarray}
\langle \Delta\hat q^2(t_f)\rangle 
=\delta^2\left(1+\frac{t_f^2\hbar^2}{4\delta^4 m^2}~\right)
=\delta^2 + \frac{\Delta^4_{SQL}}{4\delta^2 }\;,
\label{e:muwunc}
\end{eqnarray}                               
where $\langle \cdots\rangle \doteq \mbox{Tr} [\cdots \hat\varrho]$ 
denotes the ensemble average and $\Delta\hat O\doteq\hat O -
\langle \hat O\rangle $ for any operator 
$\hat O$.
Minimizing $\langle \Delta\hat q^2(t_f)\rangle $ with respect to $\delta^2$
in Eq. (\ref{e:muwunc}), one obtains $\langle \Delta\hat q^2(t_f)\rangle 
\geq\Delta^2_{SQL}$. Thus, there is a limit on the accuracy of 
a subsequent position measurement, which originates from the 
spreading of the free mass wave-function. \par
Is it possible to beat the SQL by preparing the moving mass in a different 
wave-function? The answer was given by Yuen \cite{yuen83}, who
identified a class of states---the {\em contractive} states---that have
position uncertainty that decreases versus time, and thus can go below
the SQL. Yuen also presented some measurement models of the
Arthurs-Kelly type \cite{artkel} that realize in different ways the 
state reduction toward contractive states \cite{lynch84,yuen84}. 
Ozawa subsequently proposed a different measurement scheme 
\cite{oz88} that leaves the moving mass in a contractive state, 
and can overcome the SQL. 
Moreover, after identifying a necessary condition for breaching the SQL 
\cite{oz89,oz91}, Ozawa found a class of Hamiltonians that satisfy
such condition \cite{oz90}. \par  
In this paper we present a way of engineering {\em ab initio} a
measurement scheme that beats the SQL. The scheme consists of three-steps:
a pre-squeezing, a von Neumann measurement, and a feedback. 
It turns out that our measurement is equivalent (i. e. it has the same 
outcome probability distribution and the same state reduction) to a 
model belonging to a general class previously studied by Ozawa \cite{oz90}.
The object-probe interaction of the von Neumann 
measurement is $\hat H_{I}=\hat q\hat P$. 
In our notation, lower-case operators denote system 
observables (the moving mass) and 
capital operators denote observables of the probe which, in our case, is
a single mode of the electromagnetic field. Thus $\hat q$ denotes the 
position operator, whereas the role of the linear momentum $\hat P$
for the field is played by the {\em quadrature} $\hat P\equiv\hat X_\phi
\doteq(\hat A^{\dag}e^{i\phi}+\hat Ae^{-i\phi})/2$ of the probing mode, 
with annihilation and creation operators $\hat A$ and $\hat A^{\dag}$. 
The $\hat q\hat P$ interaction may be difficult to achieve; hence 
we analyze also the case of a von Neumann 
measurement based on the interaction $\hat H_{I}=\hat q\hat 
A^{\dag}\hat A$, which is just the radiation-pressure Hamiltonian of an
interferometric measurement of a moving mirror position. 
However, in this case we show that we can only reach the SQL,
but we cannot overcome it.
\par The outline of our paper is as follows.
Section \ref{s:approx} is a brief review of the formal framework for
repeated quantum measurements. After giving the concepts of
probability operator-valued measures (POM's) and instruments, we recall the
notions of {\em precision} and {\em posterior deviation} to describe
the noise from the measurement device and the disturbance from the
state reduction. With these two concepts in mind we can recall
the precise statement of the SQL due to Ozawa \cite{oz89},
who has given a necessary condition to breach 
the SQL. In Section \ref{s:gl} we see that the SQL can be 
overcome by a Gordon-Louisell (GL) measurement \cite{gl66}, as also 
shown by Yuen \cite{yuen84,lynch84}.
After proving that every GL state reduction can be obtained by means of 
a suitable feedback mechanism, we show how in this way it is possible to 
engineer a measurement scheme that beats the SQL. As announced, the 
scheme consists of the sequence of a pre-squeezing, a von Neumann 
measurement, and a feedback. 
In Section \ref{s:ozyu} we briefly recall the Ozawa's measurement models 
that satisfy Ozawa's condition, and show how our scheme realizes some 
measurements in this class. In Section \ref{s:mir} we analyze the case of 
a von Neumann 
measurement achieved with the interaction $\hat H_{I}=\hat q\hat 
A^{\dag}\hat A$. Section \ref{s:con} closes the paper with some concluding 
remarks.

\section{Repeated and approximated measurements}\label{s:approx}

In this Section we review the main points of the theory of repeated
quantum measurements, and we resume the formulation of the SQL
given by Ozawa \cite{oz89,oz91}, based on a necessary 
condition to breach the SQL.
We will not give a complete general treatment of the subject, but only
introduce the main concepts and notation that will be used in the
following sections: for more extensive and rigorous treatments see
Refs. \cite{oz89,oz91,darank} and references therein. 

\subsection{POM's and Instruments}\label{ss:pom}

A complete description of a quantum measurement consists of both: 
{\em i)} the probability density $p(x|\hat\varrho)dx$ of the result $x$ of the
measurement when the quantum system is in the state described by the 
density matrix $\hat\varrho$; and {\em ii)} the state reduction 
$\hat\varrho\to\hat\varrho_x$ for the system immediately after 
the measurement with
outcome $x$. Both $p(x|\hat\varrho)dx$ and $\hat\varrho_x$ can 
be expressed in terms of one map, the so-called {\em instrument} $dI(x)$, 
which is a linear map on the space of trace class operators 
$\hat\varrho\rightarrow dI(x)\hat\varrho$ given by
\begin{eqnarray}                              
p(x|\hat\varrho)dx=\mbox{Tr}[dI(x)\hat\varrho ]
\;,\label{e:statinstr1}\\
\hat\varrho\rightarrow \hat\varrho_x=\frac{dI(x)\hat\varrho}
{\mbox{Tr}[dI(x)\hat\varrho ]}\;.             
\label{e:statinstr2}
\end{eqnarray}   
Hence, the instrument $dI(x)$ gives a complete description of the
quantum measurement. Sometimes, however, one is interested only in the 
probability density $p(x|\hat\varrho)$ of the measure outcome $x$,
ignoring the state reduction: in this case
it is sufficient to know the {\em probability operator-valued measure} (POM) 
$d\hat\Pi(x)$ of the measurement, which provides the
probability distribution of the readout $x$ for any state 
$\hat\varrho$ as follows
\begin{eqnarray}
p(x|\hat\varrho)dx=\mbox{Tr}[\hat\varrho d\hat\Pi (x)]\;.
\label{e:statpom}
\end{eqnarray}
By comparing Eq. (\ref{e:statpom}) with Eq. (\ref{e:statinstr1}) one
can see that for every instrument $dI(x)$ the corresponding POM 
$d\hat\Pi (x)$ is defined through the trace-duality relation 
$\mbox{Tr}[\hat\varrho d\hat\Pi (x)]=\mbox{Tr}[dI(x)\hat\varrho ]$.
The correspondence between $dI(x)$ and $d\hat\Pi (x)$ is not 
one-to-one, because the same POM can be achieved by different
instruments $dI(x)$, namely with different state reductions.
\par Insofar we have given an abstract description of the quantum measurement,
with no mention to the physical realization of the measuring
apparatus. In order to have an output state that depends on the state
before the measurement, the measurement apparatus must involve a probe 
that interacts with the system, and later is measured to yield
information on the system.
This {\em indirect} measurement scheme is completely specified once the 
following ingredients are given: i) the unitary operator 
$\hat U$ that describes the system-probe interaction; ii) the state 
$|\varphi\rangle$ of the probe before the interaction (we restrict our
attention to the case of pure-state preparation of the probe); iii) the
observable $\hat X$ which is measured on the probe. At the end
of the system-probe interaction it is possible to consider a
subsequent measurement of a (generally different) observable $\hat Y$
on the system (with outcome $y$). Then, it can be easily shown (see, for
example, Ref. \cite{darank}) that the conditional probability density 
$p(y|x)$ of getting the result $y$ from the second measurement---being $x$ 
the result of the first one---can be written in terms of the 
Born's rule $p(y|x)dy=\langle y|\hat\varrho_x|y\rangle$ upon
defining a ``reduced state'' $\hat\varrho_x$ as in Eq. 
(\ref{e:statinstr2}), where the instrument and the POM are given by
\begin{eqnarray}
d\hat\Pi (x)&=&dx\,\hat\Omega^{\dag}(x)\hat\Omega (x) 
\;,\label{e:instpomega1}\\dI(x)\hat\varrho &=&dx\,
\hat\Omega (x) \hat\varrho \hat\Omega^{\dag}(x)
\;,\label{e:instpomega2}
\end{eqnarray}              
and the operator $\hat\Omega (x)$, which acts on the Hilbert space of
the system only, is defined by the following matrix element on the
probe Hilbert space
\begin{eqnarray}
\hat\Omega (x)\doteq\langle  x|\hat U |\varphi\rangle \;,\label{e:omegae}
\end{eqnarray}       
$|x\rangle $ being the eigenvector of the observable $\hat X$ corresponding
to eigenvalue $x$.
As regards the evolution operator $\hat U$, one can neglect, for
simplicity, the free evolution during the measurement interaction
time, and consider an {\em impulsive} interaction Hamiltonian that is
switched on only for a very short time interval $\tau$ with a very large
coupling constant $K$, such that $K\tau$ is finite. For simplicity of
notation, in the following we will implicitly include $K\tau$ 
in the definition itself of the interaction Hamiltonian.
\subsection{Precision and posterior deviation}\label{ss:eps}
In any scheme for a quantum measurement, in principle there are always
two kinds of noise: {\em i)} the quantum noise of the observable 
that is intrinsic of the quantum state $\hat\varrho$, which is given by
the variance $\langle \Delta\hat q^2\rangle $; {\em ii)} the noise due to the 
measuring apparatus, which is generally non-ideal, for example, because 
of a non-unit quantum efficiency, or as a consequence of the noise due 
to a joint measurement. This leads to an output probability distribution 
$p(x|\hat\varrho)dx$ that is broader than $\langle \Delta\hat q^2\rangle $. 
A POM $d\hat\Pi(x)$ is said to be {\em compatible} with an observable 
$\hat q$ (or $\hat q$-compatible) if it satisfies the relation 
\begin{equation}
[d\hat\Pi(x),\hat q]=0
\;,\end{equation}
namely the POM has the 
same spectral decomposition of $\hat q$
\begin{equation}
d\hat\Pi(x)=G(x,\hat q)dx
=dx\int G(x,q)|q\rangle \langle q|\,dq\;,
\end{equation} 
$G(x,q) $ playing the role of a conditional probability
density for the output $x$, given that the position of the 
system was $q$. 
\par The extrinsic {\em instrumental noise} or {\em precision} 
$\epsilon^2[\varrho]$ of the apparatus with 
$\hat q$-compatible POM $d\hat\Pi(x)$ for a 
measurement of the observable $\hat q$, estimates the broadening of the 
intrinsic noise due to the 
measurement, and is defined as follows
\begin{eqnarray}
\epsilon^2 [\hat\varrho ]\doteq
\int\int (x-q)^2 \mbox{Tr} [\hat\varrho d\hat\Pi (x)|q\rangle \langle q|]
\,dq\equiv
\int\mbox{Tr} [(x-\hat q)^2 \hat\varrho d\hat\Pi (x)]\;.\label{e:pomprec} 
\end{eqnarray}                                  
The overall noise or total uncertainty $\overline{\Delta x^2}[\hat\varrho]$ 
of the measurement is the variance of the experimental 
probability distribution, namely
\begin{eqnarray}
\overline{\Delta x^2}[\hat\varrho]\equiv\mbox{E}[\Delta x^2||\hat\varrho]
\equiv\mbox{E}[x^2||\hat\varrho]-\mbox{E}[x||\hat\varrho ]^2 \;,
\label{e:totunc}
\end{eqnarray}            
where $\mbox{E}[g(x)||\hat\varrho ]\doteq\int g(x)p(x|\hat\varrho)dx$
denotes the experimental expectation value of the function $g(x)$ of
the random outcome $x$.
The total uncertainty in Eq. (\ref{e:totunc})
can be simply written as the sum of $\epsilon^2 [\hat\varrho ]$ and 
$\langle \Delta\hat q^2\rangle $ if the POM is {\em unbiased}, 
namely $\mbox{E}[x||\hat\varrho ]\equiv
\langle \hat q\rangle $ for every state $\hat\varrho$. 
In that case one has 
\begin{equation}
\langle \hat q\rangle \doteq
\mbox{Tr}[\hat q\hat\varrho]\equiv \int dx\, x \,p(x|\hat\varrho)
\;,\end{equation} 
and hence $\hat q$ can be spectrally decomposed in terms of the POM 
itself, namely
\begin{eqnarray}
\hat q= \int x d\hat\Pi (x) \Longleftrightarrow
\mbox{E}[x||\hat\varrho ]\equiv\langle \hat q\rangle  \;,
\label{e:pomunbias1}
\end{eqnarray} 
for all states $\hat\varrho$ with $\overline{\Delta x^2}[\hat\varrho]
<\infty$. The precision of a measurement realized with a
$\hat q$-compatible POM is equal to zero if and only if $G(x,q)=
\delta(x-q)$, namely if the measurement apparatus is {\em noiseless}. 
Hence
\begin{eqnarray}
\epsilon^2[\hat\varrho]=0 \qquad\Longleftrightarrow \qquad
d\hat\Pi (x)=dx|x\rangle \langle x|\label{e:noiseless}\;. 
\end{eqnarray}
In this case the total uncertainty of the measurement is just the 
variance of the system state.
\par For a $\hat q$-compatible and unbiased POM, Eq. (\ref{e:totunc}) 
can be rewritten as 
\begin{eqnarray}
\overline{\Delta x^2}[\hat\varrho ] = \epsilon^2 [\hat\varrho ]
+\langle \Delta\hat q^2\rangle  \;,
\label{e:uncunb}
\end{eqnarray}    
namely the intrinsic and the instrumental noises behave
additively. The precision $\epsilon^2 [\hat\varrho ]$ of the measurement
characterizes only the POM $d\hat\Pi(x)$, namely the probability
distribution of the measurement. On the other hand, the noise after the 
state reduction can be quantified by the {\em posterior deviation} 
$\sigma^2[\hat\varrho]$ of the instrument $I$, which is
defined as follows
\begin{eqnarray}
\sigma^2 [\hat\varrho ] \doteq
\int\int (q-x)^2 \langle q|\hat\varrho _x |q\rangle p(x|\hat\varrho)\,dx\,dq
\equiv
\int\mbox{Tr}[(\hat q -x)^2 dI(x)\hat\varrho ] \;.\label{e:instrprec} 
\end{eqnarray}                             
In Refs. \cite{oz89} and \cite{oz91} the quantity 
$\sigma^2[\hat\varrho]$ is named ``resolution'': here we suggest to adopt
the nomenclature ``posterior deviation'', as it is clearly 
a property of the reduced state $\hat\varrho_x$. 
For $\hat\varrho$ satisfying 
$\overline{\Delta x^2}[\hat\varrho ] <\infty$ and 
$\mbox{Tr}\left[\Delta\hat q^2 \int dI(x)\hat\varrho \right ]<\infty$ 
one has \cite{oz91}
\begin{eqnarray}
\sigma^2 [\hat\varrho ]= 
\int\langle \Delta\hat q^2\rangle _x p(x|\hat\varrho)dx
+\int \left[\langle \hat q\rangle _x -x\right]^2 p(x|\hat\varrho)dx
\;,\end{eqnarray}                                                          
where $\langle \cdots\rangle _x \doteq \mbox{Tr}
[\cdots \hat\varrho_x]$ denotes the conditional expectation. 
Therefore for {\em unbiased reduction}---i.e. the average value of
$\hat q$ after the state reduction is still equal to the outcome $x$
of the measurement---$\sigma^2[\hat\varrho]$ is just the variance of the
reduced state averaged over all the readouts $x$, namely  
\begin{eqnarray}
\mbox{Tr}{[\hat q\hat\varrho_x]} =x\qquad\Longrightarrow\qquad
\sigma^2 [\hat\varrho ] = 
\int\langle \Delta\hat q^2\rangle _x p(x|\hat\varrho)dx\;.
\label{e:pomunbias2}
\end{eqnarray}                                                          
\subsection{Generalized Standard Quantum Limit}
\label{ss:gsql}
Let us consider a $\hat q$-compatible and unbiased POM. 
At $t=0$ we measure the position of the system and we get
outcome $x$ and state reduction $\hat\varrho_x$, with probability 
distribution $p(x|\hat\varrho)dx$. Then we let the system evolve freely for a 
time interval $t_f$ and perform a second measurement on 
$\hat\varrho_x(t_f)$. According to a mean-value strategy, one predicts 
the second measurement to have the outcome 
$h(x)\doteq\mbox{Tr}[\hat\varrho_x(t_f)\hat q]$ with an uncertainty given by 
\begin{eqnarray}
\Delta (t_f,\hat\varrho,x)=
\left[\int [x'-h(x)]^2 p(x'|\hat\varrho_x(t_f))dx'\right] ^{1/2}\;,
\label{e:prediction}
\end{eqnarray}
which, for an unbiased POM is just given by
\begin{eqnarray}
\Delta^2 (t_f,\hat\varrho,x)=
\overline{\Delta x^2}[\hat\varrho _x (t_f)]\;.
\label{e:unbprediction}
\end{eqnarray}
The ``predictive uncertainty'' $\Delta^2(t_f,\hat\varrho)$ of the 
repeated measurement \cite{oz89} for prior state $\hat\varrho$ is 
defined as the average of $\Delta^2 (t_f,\hat\varrho,x)$ over all the outcomes 
$x$ at $t=0$, namely
\begin{equation}
\Delta^2(t_f,\hat\varrho)\doteq \int dx\,\Delta ^2(t_f,\hat\varrho,x)\,
p(x|\hat\varrho)
\;.\end{equation}
Ozawa \cite{oz91} introduced a precise definition of the
SQL as the lower bound of the predictive uncertainty 
$\Delta^2(t_f,\hat\varrho)$ with the hypothesis that the averaged
precision of the evolved state---the precision of a second measurement on the 
reduced state at time $t_f$---is greater than the posterior deviation.
Hence, for unbiased measurements satisfying the inequality
\begin{eqnarray}
\int dx\, p(x|\hat\varrho)\, \epsilon^2 [\hat\varrho_x(t_f)] \geq
\sigma^2 [\hat\varrho ]\;,                  
\label{e:ONC}
\end{eqnarray}                                     
Ozawa proved the bound 
\begin{eqnarray}
\Delta^2(t_f,\hat\varrho)\geq
|\mbox{Tr}
\left[\hat\varrho_R[\hat q(0),\hat q(t_f)]\right]|^2
\doteq\Delta^2_{SQL}
\label{e:OSQL}\;,
\end{eqnarray}  
where for the free-mass evolution
$\Delta^2_{SQL}=\hbar t_f/m$. 
Eq. (\ref{e:OSQL}) is the general version of the SQL due to Ozawa. 
Using Eq. (\ref{e:unbprediction}) one has
\begin{eqnarray}
\Delta^2(t_f,\hat\varrho)=
\int dx\, p(x|\hat\varrho)\, \epsilon^2 [\hat\varrho_x(t_f)]+
\int dx\, p(x|\hat\varrho)\,
\mbox{Tr}[\Delta\hat q^2 \hat\varrho_x(t_f)]\rangle \geq
\Delta^2_{SQL}\label{e:sql1}\;.
\end{eqnarray}                
Therefore, a {\em necessary} condition for breaching the SQL corresponds
to negating the hypothesis (\ref{e:ONC}), namely that the precision of 
the measurement is less than the posterior deviation. We will call this
condition ONC (Ozawa necessary condition).
We recall that, while the precision (\ref{e:pomprec}) is a measure of the 
added noise and is related to the POM, the posterior deviation 
(\ref{e:instrprec}) is related to the state reduction. As an example, it 
is interesting to evaluate both $\sigma^2 [\hat\varrho ]$
and $\epsilon^2 [\hat\varrho]$ for a generalized von Neumann measurement model
with interaction Hamiltonian of the form $\hat H_{I}=\hat q
\hat O$, where $\hat O$ is a generic observable of the probe, and 
$\hat q$ is just the quantity of the system that we want to measure---the 
position in our case (in the ``standard'' von Neumann model \cite{vn55}
$\hat O\equiv\hat P$, where $\hat P$ is the linear momentum of the probe). From
Eqs. (\ref{e:instpomega1}), (\ref{e:instpomega2}) and
(\ref{e:omegae}), it is apparent that both the POM and the instrument are 
only functions of the operator $\hat q$, so that the precision 
(\ref{e:pomprec}) and the posterior deviation (\ref{e:instrprec}) coincide.
This means that condition (\ref{e:ONC}) is verified only with the equal
sign, and the von Neumann model just achieves the SQL.

\section{Evolution operator for beating the SQL}\label{s:gl}

In this Section we will resume the possibility to overcome the SQL 
by reducing the system to a contractive state. 
Then we will design an 
evolution operator $\hat U$ that realizes such reduction. 
Finally, we will generalize $\hat U$ to a class of
operators which fulfill ONC.

\subsection{The contractive states}\label{ss:cs}

The original argument for the validity of the SQL due to
Braginsk\u{\i}i and Caves \cite{bragi74,caves80}, was the following. 
Consider a mass $m$, whose position $\hat q$ has been 
measured once within a certain precision, and let it freely evolve for a time 
$t_f$. In the Heisenberg picture the position operator evolves {\em 
classically}, as $\hat q(t_f) =\hat q(0)+\hat p(0)t_f/m$. From the
Heisenberg uncertainty principle, one obtains the following position 
variance constraint
\begin{eqnarray}
\langle \Delta\hat q^2(t_f)\rangle &=&
\langle \Delta\hat q^2(0)\rangle +\langle \Delta\hat p^2(0)\rangle 
\left(\frac{t_f}{m}\right)^2\nonumber\\ &\geq&
\langle \Delta\hat q^2(0)\rangle +\frac{\hbar^2}
{4\langle \Delta\hat q^2(0)\rangle }
\left(\frac{t_f}{m}\right)^2\geq \Delta^2_{SQL}
\;.\label{crms}
\end{eqnarray}
However, as pointed out by Yuen \cite{yuen83,yuen84}, this is not the 
correct general expression for $\langle \Delta\hat q^2(t_f)\rangle $, 
because it neglects 
the correlation term 
$2\mbox{Re}\langle \Delta\hat q(0)\Delta\hat p(0)\rangle
\doteq\langle \Delta\hat q(0)\Delta\hat p(0) +
\Delta\hat p(0)\Delta\hat q(0)\rangle  $, 
hence implicitly assuming that it is greater than or equal to zero. 
The complete expression for the position uncertainty is
\begin{eqnarray}
\langle \Delta\hat q^2(t_f)\rangle =\langle \Delta\hat q^2 (0)\rangle  + 
2\mbox{Re}\langle \Delta\hat q(0)\Delta\hat p(0)\rangle \frac{t_f}{m} +
\langle \Delta\hat p^2 (0)\rangle \left(\frac{t_f}{m}\right)^2
\;.\label{e:yrms}
\end{eqnarray}  
As a matter of fact, there is a class of states, the {\em contractive 
states} (CS), which have a negative correlation term in Eq. 
(\ref{e:yrms}). 
When the reduced state $\hat\varrho_x$ for the free mass after the first
measurement is a CS, $\langle \Delta\hat q^2(t_f)\rangle _x$ decreases in time
before reaching a minimum value. If the measurement is sufficiently 
precise, the SQL can be overcome. In fact, 
if $\langle \Delta\hat q^2(t_f)\rangle _x
\leq\langle \Delta\hat q^2 (0)\rangle _x $ for every outcome $x$, 
then this is true also on the average, namely
\begin{eqnarray}
\int dx\, p(x|\hat\varrho)\,
\langle \Delta\hat q^2 (t_f)\rangle _x\leq
\int dx\, p(x|\hat\varrho)\,\langle \Delta\hat q^2 (0)\rangle _x
\;.\label{yrms2}
\end{eqnarray}
Eqs. (\ref{e:uncunb}) and (\ref{e:unbprediction}) leads to
\begin{eqnarray}
\Delta^2(t_f,\hat\varrho)=
\int dx\, p(x|\hat\varrho)\, \epsilon^2 [\hat\varrho_x(t_f)]+
\int dx\, p(x|\hat\varrho)\,
\langle \Delta\hat q^2 (t_f)\rangle _x
\label{e:usql}\;,
\end{eqnarray}  
which, for noiseless measurements $\epsilon^2 [\hat\varrho_x(t_f)]=0$,
gives 
\begin{eqnarray}
\Delta^2(t_f,\hat\varrho)=
\int dx\, p(x|\hat\varrho)\, \langle \Delta\hat q^2 (t_f)\rangle _x \leq
\int dx\, p(x|\hat\varrho)\, \langle \Delta\hat q^2 (0)\rangle _x 
\label{e:usql2}\;.
\end{eqnarray}  
A particular example of a CS are the squeezed states, or 
{\em twisted coherent state} (TCS)\cite{yuen76}. A TCS 
$|\mu\nu\alpha\omega\rangle $ 
for the mass has the following position representation
\begin{eqnarray}
\!\!\langle q|\mu\nu\alpha\omega\rangle =\left(
\frac{mw}{\pi\hbar|\mu -\nu|^2} 
\right)^{\frac14}
 \exp\left[
-\frac{mw}{2\hbar}\frac{1+2i\xi}{|\mu -\nu|^2}(q-q_0)^2 + 
\frac{i}{\hbar}p_0 (q-q_0)
\right],\label{e:tcs}
\end{eqnarray}                       
with fluctuations
\begin{eqnarray}
\langle \Delta\hat
q^2(t_f)\rangle =\frac{\hbar}{2m\omega}|\mu-\nu|^2-\frac{2\hbar\xi}{m}t_f+
\frac{\hbar\omega}{2m}|\mu+\nu|^2t_f^2
\label{e:deltcs}\;,
\end{eqnarray}
where $|\mu|^2-|\nu|^2=1$, $\xi=\mbox{Im}(\mu^*\nu)>0$ and $\alpha=q_0+ip_0$, 
$q_0$ and $p_0$ being real.
The correlation function $\mbox{Re}\langle 
\Delta\hat q(0)\Delta\hat p(0)\rangle =
-\xi\hbar$, is negative for $\xi >0$, so that the position uncertainty 
$\langle \Delta\hat q^2(t_f)\rangle $ decreases in time till, at 
$t_{\scriptscriptstyle M}
=2\xi/\omega|\mu+\nu|^2$ it reaches the minimum value 
$\langle \Delta\hat q^2(t_{\scriptscriptstyle M})\rangle =(4\xi)^{-1}
(\hbar t_{\scriptscriptstyle M}/m)<\Delta^2_{SQL}$ for sufficiently large 
$\xi$ \cite{oz89}. This implies that, for noiseless 
detection [Eq. (\ref{e:noiseless})] the total uncertainty 
of the measurement goes below the SQL. A detection scheme of 
this type is described by the measurement with 
state reduction $\hat\varrho\rightarrow
\hat\varrho_Q=|\mu\nu Q\omega\rangle  \langle \mu\nu Q\omega|$
and probability distribution
$p(Q|\hat\varrho)=\langle  Q|\hat\varrho|Q\rangle $, $Q$ being
the output of a single measurement. This is a GL measurement with
reduction operator $\hat\Omega$ given by
\begin{eqnarray}
\hat\Omega(Q) =|\mu\nu Q\omega\rangle \langle  Q|\;.\label{e:gl}
\end{eqnarray}  
It is easily proved that this scheme satisfies the request of unbiasedness 
for both the POM and the state reduction, and, as expected, Eq.
(\ref{e:ONC}) is contradicted, because 
$\sigma^2 [\hat\varrho ]=
\int dQ p(Q|\hat\varrho)\mbox{Tr}[\Delta\hat q^2\hat\varrho_Q(0)]>
\int dQ p(Q|\hat\varrho)\epsilon^2[\hat\varrho_Q(t_f)]\equiv 0$.
\par Yuen \cite{lynch84,yuen84} and Ozawa have given different interaction
Hamiltonians that realize the operator $\hat\Omega(Q)$ in Eq. (\ref{e:gl}). 
In the next Section, we will engineer {\em ab initio} a measurement 
scheme that has the state reduction $|\mu\nu Q\omega\rangle $. 

\subsection{Gordon-Louisell measurements}\label{ss:glm}

The definition of a general GL measurement is given by a state reduction 
operator of the form
\begin{eqnarray}
\hat\Omega(x)=|\psi_x\rangle \langle \theta_x|\;,\label{e:mauro1}
\end{eqnarray}
where $|\theta_x\rangle $ is a complete set, i. e. 
$\int dx|\theta_x\rangle \langle \theta_x|=\hat 1$
and $|\psi_x\rangle $ is a normalized physical state.
The operator $\hat\Omega(x)$ in Eq. (\ref{e:mauro1}) 
abstractly represents an indirect measurement that leaves 
the system in the state $|\psi_x\rangle $, independently on the input state, 
and has probability distribution for the outcome $x$ given by 
$p(x|\hat\varrho)=\langle \theta_x|\hat\varrho|\theta_x\rangle $. In 
the particular case that ${|\theta_x\rangle }$ is also an orthogonal set
$\langle \theta_x|\theta_x^{\prime}\rangle =\delta (x-x^{\prime})$, there is a 
unitary operator $\hat U$ on the system-probe Hilbert space 
${\cal H}_S\otimes{\cal H}_P$ that gives $\hat\Omega(x)$ in Eq. 
(\ref{e:mauro1}) through Eq. (\ref{e:omegae}). In fact, let us
define the self-adjoint operator $\hat\theta$ such that 
$\hat\theta|\theta_x\rangle =\theta_x|\theta_x\rangle $. Then, let us 
consider an indirect measurement scheme for $\hat\theta$, 
with the probe prepared in the (pure) state 
$|\varphi\rangle $. Comparing Eq. (\ref{e:omegae}) with Eq. (\ref{e:mauro1}), 
we see that $\hat U$ must satisfy the identity
\begin{eqnarray}
\langle \theta_y|\otimes\langle \theta_x|\hat U|\theta_z\rangle \otimes
|\varphi\rangle =\langle \theta_y|\psi_x\rangle \langle \theta_x|\theta_z
\rangle =\langle \theta_y|\psi_x\rangle \delta(x-z)\;.\label{e:mauro3}
\end{eqnarray}
Hence, a unitary operator $\hat U$ that has the matrix elements 
(\ref{e:mauro3}) can be chosen as achieving the following linear 
transformation
on ${\cal H}_S\otimes{\cal H}_P$
\begin{eqnarray}
\hat U|\theta_x\rangle \otimes|\varphi\rangle =|\psi_x\rangle \otimes|
\theta_x\rangle \;.\label{e:mauro4}
\end{eqnarray}
How can we design such an operator $\hat U$? 
Consider the operator $\hat R$ that effects a rotation by $\pi /2$ in 
the $(q,Q)$ plane, namely
\begin{eqnarray}
\hat R=\exp\left[\frac{i\pi}{2\hbar}(\hat p\hat Q-\hat q\hat P)\right]\;.
\end{eqnarray}
Apart from an inversion and a trivial overall phase factor, 
the operator $R$ corresponds to a mode-permutation operator since one has
\begin{eqnarray}
\hat R \psi_1(q)\psi_2(Q)=\psi_2(-q)\psi_1(Q)
\;. \label{e:mauro5}
\end{eqnarray}
Then we introduce the notion of feedback operator $\hat F(x)\in{\cal H}_S$, 
namely a self-adjoint operator that parametrically depends on the
measure outcome $x$. This can be any self-adjoint operator function of
$x$: however, we need to restrict the class of
operator functions $\hat F(x)$ such that the integral 
$\hat F_S(\hat X)\doteq\int dx\,\hat F(x)
\otimes|\theta_x\rangle \langle \theta_x|$ 
converges to a well defined self-adjoint operator acting on 
${\cal H}_S\otimes{\cal H}_P$, with 
$\hat X|\theta_x\rangle =x|\theta_x\rangle $ \cite{nota}. 
We choose the feedback Hamiltonian such that it connects
the vectors $|\varphi\rangle $ and $|\psi_x\rangle $, namely 
\begin{eqnarray}
|\psi_x\rangle \doteq \exp\left[-\frac{i}{\hbar}\hat F(x)\right]
|\varphi\rangle \;.
\end{eqnarray}
Finally, we write the operator $\hat U$ as follows
\begin{eqnarray}
\hat U=\exp\left[-{i\over\hbar}\hat F_S(\hat X)\right]
\exp\left[\frac{i\pi}{2\hbar}(\hat p\hat Q-\hat q\hat
P)\right]\;,\label{e:mauro6}
\end{eqnarray}                                          
\par Now we specialize Eq. (\ref{e:mauro6}) to the case described by Eq.
(\ref{e:gl}). 
The wave-functions $|\psi_x\rangle $ and $|\theta_x\rangle $ correspond
to $|\mu\nu Q\omega\rangle $ and $|Q\rangle $ respectively, whereas the
probe observable $\hat\theta$ is the position $\hat Q$, 
with outcome $Q$. This implies that the feedback operator has to 
shift the position of the system by a quantity equal to the output
$Q$, namely $\hat F_S(\hat Q)=\hat p\hat Q$. 
Hence, the evolution operator has the form
\begin{eqnarray}
\hat U=\exp\left[-\frac{i}{\hbar}\hat p\hat Q\right]
\exp\left[\frac{i\pi}{2\hbar}(\hat p\hat Q-\hat q\hat P)\right]\;.
\label{e:mauro7}
\end{eqnarray}                                          
We emphasize that, as $\hat U$ contains a permutation operator, 
the initial probe state must be chosen to be a TCS. 
Hence, if the initial state of the system is $|\psi\rangle $, we get 
$\hat R|\psi\rangle \otimes|\mu\nu\alpha\omega\rangle \propto  
|\mu\nu(-\alpha)\omega\rangle \otimes|\psi\rangle $: 
in practice, the position of
the object is indirectly squeezed by squeezing the probe position, and
then exchanging the state of the system with that of the probe. 
For simplicity, we choose $\alpha =0$, which means that the initial 
position of the probe is equal to zero. After that, the feedback operator 
shifts the position of the system so that it finally corresponds to the 
output of the measurement, that is 
$\exp[-\frac{i}{\hbar}\hat F(Q)]|\mu\nu 0 \omega\rangle =
|\mu\nu Q\omega\rangle $. 
Now we relax the value of the coupling constants in the interaction 
Hamiltonians and rewrite the evolution operator (\ref{e:mauro7}) 
in the general form
\begin{eqnarray}
\hat U=\exp\left[-\frac{i}{\hbar}g_1\hat p\hat Q\right]
\exp\left[\frac{i\pi}{2\hbar}g_2(s^{-1}\hat p\hat Q-s\hat q\hat P)\right]\;.
\label{e:mauro8}
\end{eqnarray}                                          
As regards the concrete feasibility of the evolution described by Eq. 
(\ref{e:mauro8}), the feedback part can be achieved by a position transducer,
which displaces the mass by an amount $g_1 Q$ for every position outcome
$Q$. 
Notice that we do not need to realize the
feedback Hamiltonian $g_1\hat p\hat Q$, because as a consequence of 
Eq. (\ref{e:omegae}) this is equivalent to the system Hamiltonian 
$g_1\hat p Q$, where the coupling is rescaled by the 
eigenvalue $Q$ in place of the operator $\hat Q$. 
The permutation operator will be discussed in the following section.
\section{Comparison with previous models and realization}\label{s:ozyu}
In the first part of this Section we compare the unitary evolutions of the 
Ozawa's measurement models \cite{oz90} satisfying ONC with 
our operator in Eq. (\ref{e:mauro8}). In the second part we factorize
our evolution into three steps, expressing the permutation operator by
means of a pre-squeezing of both system and probe, followed by a von Neumann
interaction $\hat H_I=\hat q \hat P$.

\subsection{Ozawa's Hamiltonians and the conditions for beating the SQL} 
\label{ss:oz}                                                              

In Ref. \cite{oz90} Ozawa gives a class of measurement models
that satisfy ONC. He explicitly writes the unitary system-probe 
evolution in an impulsive regime in terms of an interaction Hamiltonian 
$\hat H_I$, generalizing the standard von Neumann measurement model with 
$\hat H_I=\hat q\hat P$ \cite{vn55}. The Hamiltonian is given by
\begin{eqnarray}
\hat H_I= -k_+ \hat Q\hat p -k_- \hat q\hat P 
+ k_z (\hat q\hat p-\hat Q\hat P)\;,
\label{e:hint}
\end{eqnarray}                                         
with $k_+\;, k_-$ and $k_z$ as real parameters. For probe initial state 
$|\varphi\rangle $, with $\varphi (Q)=\varphi(-Q)$ and 
$\langle \varphi |\hat Q|\varphi\rangle =0$ (which is
equivalent to condition $\alpha=0$ in the previous section),
and for the system initially in the state 
$|\psi\rangle $, Eqs. (\ref{e:pomprec}) and (\ref{e:instrprec}) become
\begin{eqnarray}
\epsilon^2 [\psi]&=&(1-c)^2\langle\psi |\hat q^2 |\psi \rangle +
d^2\langle \varphi|\hat Q^2|\varphi\rangle  \;,\label{e:ozprec1}\\
\sigma^2[\psi]&=&(a-c)^2\langle\psi |\hat q^2 |\psi \rangle +
(b-d)^2\langle \varphi|\hat Q^2|\varphi\rangle \;. 
\label{e:ozprec2}
\end{eqnarray}    
Here the coefficients of the dynamics $(a,b,c,d)$---grouped in a
matrix form that will be used in the following---are
\begin{eqnarray}
\left[\begin{array}{cc} a &-b\\-c& d\end{array}\right]=     
\left[\begin{array}{lc} 
\cosh {\cal K} +k_z \frac{\sinh{\cal K} }{{\cal K}} &
k_+ \frac{\sinh{\cal K}}{{\cal K}} \\
k_- \frac{\sinh{\cal K}}{{\cal K}} &
\cosh{\cal K} -k_z \frac{\sinh{\cal K}}{{\cal K}}\end{array}\right]    
\label{e:abcd}
\end{eqnarray}                          
where ${\cal K}=\sqrt{k_z^2 +k_+k_-}$ can be either real or pure
imaginary. 
\par 
The degrees of freedom of the above equations can be reduced by applying
the criteria for a plausible object-probe interaction: one requests that
if the uncertainty of the prior probe coordinate $\langle \varphi|\Delta
\hat Q^2|\varphi\rangle $ tends to zero, then the precision and the posterior
deviation tend to zero. From Eqs. (\ref{e:ozprec1}) and
(\ref{e:ozprec2}) it follows that one needs 
$a=c=1$. It can be shown \cite{oz90} that the condition $c=1$ is also a
consequence of the unbiasedness of the POM. Moreover, as $ad-bc=1$, it 
turns out that $b=d-1$, and using Eqs. (\ref{e:abcd}) one has
$d=2\cosh{\cal K}-1$. With the above restrictions, the precision and
the posterior deviation become 
$\epsilon^2=d^2\langle \varphi|\Delta\hat Q^2|\varphi\rangle $ and 
$\sigma^2 =\langle \varphi|\Delta\hat Q^2|\varphi\rangle $, independently 
on the initial system state. Thus, the ONC for beating the SQL can be
expressed as $|d|<1$. In summary, we have
\begin{eqnarray}
&&a=1\;,\quad b=d-1\;,\quad c=1 \;,\nonumber\\
&&\label{e:under}\\
&&|d|=|2\cosh{\cal K} -1|<1 \;,
\quad\mbox{ONC}\Longleftrightarrow |d|<1\;.\nonumber
\end{eqnarray}                      
If ${\cal K} $ is real, then $d>1$; but, if ${\cal K}$ is pure imaginary, 
$d=2 \cos |{\cal K} |-1$, and $-3<d<1$. Thus, Ozawa's Hamiltonians satisfy 
ONC for certain values of the parameters $(k_+,k_-,k_z)$ which can be 
found by comparing Eqs. (\ref{e:under}) with Eqs. (\ref{e:abcd}), namely
\begin{eqnarray}
k_+ = 2k_z \;,\qquad \cosh{\cal K} = 1 + \frac{k_z}{k_-} \;,
\qquad -1 < \frac{k_z}{k_-} <0  \;.
\label{e:klink}
\end{eqnarray}                                                        
In particular, among the Hamiltonians (\ref{e:hint}), the model described 
by Ozawa in reference \cite{oz88}, for $k_z=\pi/3\sqrt{3}$ and 
$k_{\pm}=\pm2k_z$, realizes the GL scheme
$|\mu\nu Q\omega\rangle \langle  Q|$ with $d=0$.
On the other hand, the original von Neumann model \cite{vn55} corresponds 
to $d=1$ and $\epsilon^2=\sigma^2=\langle \varphi|\Delta\hat Q^2|
\varphi\rangle $. This means that there is a continuum of models for
$|d|<1$ that have a better precision than 
$\sigma^2=\langle \varphi|\Delta\hat Q^2|\varphi\rangle $, and can 
possibly circumvent the SQL.
\par We now look for the relation between our model and the model just
described. We use the following realization of the angular momentum
(complex) Lie algebra $gl(2,C)$ 
\begin{eqnarray}
\hat J_+\doteq\frac{i\hat Q\hat p}{\hbar}\;,\qquad
\hat J_-\doteq\frac{i\hat q\hat P}{\hbar}\;,\qquad
\hat J_z\doteq\frac{i}{2\hbar}(\hat Q\hat P-\hat q\hat p)\;.     
\label{e:su2}
\end{eqnarray}                                                  
One can easily verify that the above operators satisfy the $gl(2,C)$
commutation relations $[\hat J_+,\hat J_-]=2\hat J_z$ and $[\hat J_z,\hat
J_{\pm}]=\pm \hat J_{\pm}$, so that the exponential of their linear
combinations can be faithfully represented by $2\times2$ Pauli
matrices, independently of the value of the angular momentum $J$ [the 
group $gl(2,C)$ is complex, and the fact that the realization 
(\ref{e:su2}) does not preserve Hermitian 
conjugation is irrelevant for the group multiplication law]. 
Notice that the matrix in Eq. (\ref{e:abcd}) 
is nothing but the Pauli representation of the evolution
operator $\hat U=\exp(-i\hat H_I/\hbar)$ in terms of the $gl(2,C)$ 
algebra realization (\ref{e:su2}). 
Both our operator $\hat U$ in Eq. (\ref{e:mauro7}) and
Ozawa's in Eq. (\ref{e:abcd}) for  $k_z=\pi/3\sqrt{3}$ and 
$k_{\pm}=\pm2k_z$, lead to the same coefficients $(a,b,c,d)$, namely
$(1,-1,1,0)$. In fact, we have just seen that they both realize
$\hat\Omega(Q)$ in Eq. (\ref{e:gl}).
In the general case, the relation between the coefficients of the
operator (\ref{e:mauro8}) and the ones in Eq. (\ref{e:abcd}) are
\begin{eqnarray}
a&=&\cos(\frac{\pi g_2}{2}) + g_1 s\sin(\frac{\pi g_2}{2}) \;,\qquad
b=-\frac{1}{s}\sin(\frac{\pi g_2}{2})+g_1\cos(\frac{\pi g_2}{2})\;,\nonumber\\
&&\label{e:bchgl}\\
c&=& s \sin(\frac{\pi g_2}{2})\;,\qquad d=\cos(\frac{\pi g_2}{2})\;.\nonumber 
\end{eqnarray}    
The last of Eqs. (\ref{e:bchgl}) clearly states that $d$ belongs to
the interval $[-1,1]$, namely all possible models realizing the
condition $\epsilon <\sigma $ are included.

\subsection{Realization through $GL(2,C)$ elements} 
\label{ss:su2}                                                              

In this Subsection we suggest a measurement scheme that realizes the
evolution operator in Eq. (\ref{e:mauro8}) or Eq. (\ref{e:abcd}) through 
three steps in the $GL(2,C)$ group.
We are describing an indirect measurement of the system 
position $\hat q$ through detection of the probe position $\hat Q$,
and the central element of our scheme is the von Neumann Hamiltonian 
$\hat J_-\doteq i\hat q\hat
P/\hbar$, which entangles the system object with the probe by shifting 
$\hat Q$ by $\hat q$. Moreover, Eqs.
(\ref{e:mauro8}) and (\ref{e:su2}), suggest that {\em after} any 
detection with output $Q$, a feedback of the form 
$\hat J_+\doteq i\hat Q\hat p/\hbar$ is requested. This implies that, 
among the six evolution operators that can be obtained by permuting the
exponentials of $(\hat J_+,\hat J_-,\hat J_z)$, we will not consider the 
three permutations with $\exp\left(\zeta_+ \hat J_+\right)$ applied before 
$\exp\left(\zeta_- \hat J_-\right)$: 
in the next Subsection we will analyze the remaining three cases.

\subsection{Feedback-assisted measurement}
\label{ss:schemeIII}   

Consider the sequence
\begin{eqnarray}
\exp\left(\zeta_+ \hat J_+\right)\exp\left(\zeta_- \hat J_-\right)
\exp\left( 2\zeta_z \hat J_z\right) = 
\exp\left( k_+\hat J_+ + k_-\hat J_- +2k_z\hat J_z\right)\;.
\label{e:schemeIII}
\end{eqnarray}  
In this model, the operator $\exp\left( 2\zeta_z \hat J_z\right)=
\exp{\left[i\zeta_z (\hat Q\hat P-\hat q\hat p)/\hbar\right]}$ 
does not entangle the system with the probe, but just {\em pre-squeezes}
the states of both. Then, $\exp\left(\zeta_- \hat J_-\right)=
\exp{\left(i\zeta_-\hat q\hat P/\hbar\right)}$ entangles the probe with
the system. Finally, the operator $\exp\left(\zeta_+ \hat J_+\right)=
\exp{\left(i\zeta_+ \hat Q\hat p/\hbar\right)}$
corresponds to a feedback mechanism that shifts the system position by
$\zeta_+Q$. By means of the Lie algebra decomposition formulas \cite{ban}, 
the relations between the parameters $\zeta_+, \zeta_-, \zeta_z$
and the Ozawa's write as follows
\begin{eqnarray}
a=(1+ \zeta_+ \zeta_- )e^{\zeta_z}\;,\qquad -b =\zeta_+ e^{-\zeta_z}
\;,\qquad -c=\zeta_- e^{\zeta_z}\;,\qquad d=e^{-\zeta_z}
\;.\label{e:bchIIIabcd}
\end{eqnarray}        
The range of the coefficients $(\zeta_+, \zeta_-, \zeta_z )$ in which
the SQL can be overcome is [see Eq. (\ref{e:under})]
\begin{eqnarray}
\zeta_-= -e^{-\zeta_z}\;,\qquad\zeta_+ = e^{\zeta_z}-1\;,
\qquad\zeta_z >0\;.
\label{e:underIII}
\end{eqnarray}      
Now a problem arises, regarding the class of Ozawa's Hamiltonians
represented by this model. In fact, as $d=e^{-\zeta_z}>0$, it is clear
that a part of the evolutions that circumvent the SQL are excluded by
this parameterization. In particular, rewriting $(k_+, k_-, k_z)$ in 
terms of $(\zeta_+,\zeta_-,\zeta_z)$ as in (\ref{e:underIII}), one gets
\begin{eqnarray}
k_z &=& \left | \frac{e^{-\zeta_z}-1}{e^{-\zeta_z}+3} \right |^{1/2}
\arcsin \left [ \frac{1-e^{-\zeta_z}}{2}
\left | \frac{e^{-\zeta_z}+3}{e^{-\zeta_z}-1} \right |^{1/2} \right ]
\;,\nonumber\\ 
&&\label{e:klinkIII}\\
k_+ &=& 2k_z\;, \qquad k_- = \frac{2k_z}{e^{-\zeta_z}-1}\;,
\quad -\frac{1}{2} < \frac{k_z}{k_-} <0  \;,\nonumber
\end{eqnarray}
where a narrower interval for $k_z/k_-$ than in Eq. (\ref{e:klink}) is
obtained. One is lead to think that there must be some other schemes
that give the Ozawa's Hamiltonian for $d>0$. However, an outlook 
at the remaining two permutations with 
$\exp\left(\zeta_- \hat J_-\right)$ acting before 
$\exp\left(\zeta_+ \hat J_+\right)$ reveals that also in these cases
$d=e^{-\zeta_z}>0$, which implies that a substantial part of the Ozawa's
Hamiltonians cannot be realized with our scheme. 
Notice, however, that the Hamiltonian with coefficients 
$k_z=\pi/3\sqrt{3}\;,k_+= 2k_z\;,k_-=-2k_z$, which beats the SQL, is
included in our scheme (\ref{e:schemeIII}), and can be achieved in the 
limit of infinite $\zeta_z$, that is for very high squeezing for both probe 
and object preparations. The realization of the squeezing of a 
mass position remains as a challenge for experimentalists.
We stress the fact that, if either $\zeta_z$ or $\zeta_+$ in Eq. 
(\ref{e:schemeIII}) are set equal to zero and the conditions for 
breaching the SQL are imposed, the remaining two coefficients 
become equal to zero, too. That is, both the feedback and the dilatation
are essential in order to achieve a detection scheme that possibly
beats the SQL.
\par Now we calculate the operator 
$\hat\Omega(Q)$ relative to the evolution operator $\hat U$ just 
described. Because of the peculiar form of the operator $\hat U$, 
also the operator $\hat{\Omega}(Q)$ can be factorized into three parts.
By separating the exponentials that act only on the system from those
acting also on the probe, we obtain
\begin{eqnarray}
\hat{\Omega}(Q)&=&\exp{\left(i\zeta_+
Q\hat{p}/\hbar\right)}
\langle Q| \exp{\left(i\zeta_- \hat{q}\hat{P}/\hbar\right)}
\exp{\left(i\zeta_z \hat{Q}\hat{P}/\hbar\right)}| \varphi \rangle\nonumber\\ 
&\times&\exp{\left(-i\zeta_z \hat{q}\hat{p}/\hbar\right)}\;.\label{e:facto}
\end{eqnarray}
By imposing the conditions (\ref{e:underIII}) for
a plausible measurement, the operator (\ref{e:facto}) becomes
\begin{eqnarray}
\hat\Omega(Q)
=\exp{\left[i\left(e^{\zeta_z}-1\right)Q\hat{p}/\hbar\right]}  
\exp{\left(-i\zeta_z \hat{q}\hat{p}/\hbar\right)}     
\varphi [e^{\zeta_z}(Q- \hat{q})]                              
\label{e:bchome}\;.
\end{eqnarray}                                                           
In particular, for probe initial state $|\varphi\rangle $ chosen as a
TCS $|\varphi\rangle =|\mu\nu 0 \omega\rangle $, we obtain
\begin{eqnarray}
\hat\Omega(Q)&=&
\exp{\left[i\left(e^{\zeta_z}-1\right)Q\hat{p}/\hbar\right]}  
\exp{\left(-i\zeta_z \hat{q}\hat{p}/\hbar\right)} \nonumber\\ 
&&\left(\frac{mw}{\pi\hbar|\mu -\nu|^2} 
\right)^{\frac14}
\exp\left[
-\frac{mw}{2\hbar}\frac{1+2i\xi}{|\mu -\nu|^2}
[e^{\zeta_z}(Q- \hat{q})]^2 
\right]
\label{e:bchtcsome}\;.
\end{eqnarray}                       
Thus, the POM becomes 
\begin{eqnarray}
d\hat\Pi(Q)=\left(\frac{mw}{\pi\hbar|\mu -\nu|^2} 
\right)^{\frac12}
\exp\left[
-\frac{mw}{\hbar}\frac{1}{|\mu -\nu|^2}
[e^{\zeta_z}(Q-\hat q)]^2 
\right]dQ
\label{e:bchtcspom}\;,
\end{eqnarray}  
which is Gaussian and unbiased. Hence, the precision 
$\epsilon^2[\hat\varrho]$, is independent on the state $\hat\varrho$,
and is given by the variance of the Gaussian, namely
$\epsilon^2[\hat\varrho]=e^{-2\zeta_z}\langle \varphi|\Delta\hat Q^2|
\varphi\rangle $. This is precisely what we expected, because, 
as we have seen in Subsection \ref{ss:oz}, 
$\epsilon^2[\hat\varrho]=d^2\langle \varphi|\Delta\hat Q^2|\varphi\rangle $.
We recall that, for very high squeezing, the operator 
(\ref{e:bchtcsome}) reduces to the GL operator (\ref{e:gl}), and 
$\epsilon^2[\hat\varrho]$ tends to zero.

\section{Radiation-mirror interaction}\label{s:mir}

In our knowledge there is no viable way to achieve the von Neumann
interaction $\hat H\propto\hat q\hat P$. In order to approach its
behavior, other Hamiltonians have been suggested. In particular, 
Walls et al. \cite{walls93} described the interaction $\hat H_I=
\hat A^{\dag}\hat A\hat q$, where a harmonic
oscillator, the end mirror of a cavity, is interacting with the radiation 
in the cavity through radiation 
pressure. In this Section, we examine the situation in which the
mirror can be considered as a free mass, similarly to the case of a typical
gravitational waves interferometer, where the wave detector that is
attached to the mirror is a very massive bar ($m\sim 100 kg$
\cite{caves80}), which has negligible oscillation frequency. We 
rewrite the radiation interaction Hamiltonian in the form
\begin{eqnarray}
\hat{H}_I=-\hbar K_m \hat q\hat A^{\dag}\hat A\;.
\label{e:mirror}
\end{eqnarray}
For the mirror at one hand of a cavity of length $L$, the 
coupling constant $K_m$ can be derived \cite{walls93} as $K_m=\omega_0/L$, 
where $\omega_0$ is the resonance frequency of the cavity. With the 
impulsive approximation $K_m\tau\sim 1$ the unitary evolution operator
becomes simply
\begin{eqnarray}
\hat{U}=\exp(-i\hat{H}_I\tau/\hbar)=
\exp(i\underline{\hat q}\hat A^{\dag}\hat A) \;,
\label{e:evomir}
\end{eqnarray}
where the mass position $\underline{\hat q}\doteq\hat q/l_{\tau}$  
is rescaled by $l_{\tau}=L/\omega_0\tau$.
We know that the von Neumann-type measurement schemes allow only to 
reach---not to beat---the SQL, and from the previous sections we
learned that in order to overcome the SQL we need to add a
feedback and a pre-squeezing to the present scheme. Therefore,
we propose the following simple detection
scheme. The probe is prepared in a highly excited coherent state 
$|\varphi\rangle =|\alpha\rangle$, where $\alpha =-i|\alpha |$, and
then we detect (by homodyning it) the scaled quadrature 
$\hat{X}/|\alpha | 
=(\hat A^{\dag}+\hat A)/2|\alpha | $.
The operator $\hat\Omega$ in Eq. (\ref{e:omegae}) becomes
\begin{eqnarray}
\hat\Omega_{\hat q}(X)= \left( \frac{2|\alpha |^2}{\pi}\right)^{1/4} 
\exp\left[-|\alpha |^2 (X +i e^{i\underline{\hat q}})^2 
-\frac{|\alpha |^2 }{2}(1+e^{2i\underline{\hat q}})\right]  \;,
\label{e:omegamir}
\end{eqnarray}        
and for the POM we have
\begin{eqnarray}
d\hat\Pi_{\hat q}(X)&=&dX \left(\frac{2|\alpha
|^2}{\pi}\right)^{1/2}
\exp\left[-2|\alpha |^2(X-\sin\underline{\hat q})^2 \right]\;.
\label{e:pomir}
\end{eqnarray}        
For states with $\langle \underline{\hat q}\rangle\ll 1$ and
$\langle\Delta\underline{\hat q}^2\rangle\ll 1$, such that
$\sin\underline{\hat q}$ can be approximated as
$\sin\underline{\hat q}\simeq\underline{\hat q}$, one has the
unbiased POM
\begin{eqnarray}
d\hat\Pi_{\hat q}(X)&=&dX \left(\frac{2|\alpha
|^2}{\pi}\right)^{1/2}
\exp\left[-2|\alpha |^2(X-\underline{\hat q})^2 \right]\;.
\label{e:pomirsmall}
\end{eqnarray}        
Hence, for small mirror displacements, we can check if there are violations of 
the SQL. Notice that, within the small $q$ approximation, 
the operator (\ref{e:omegamir}) becomes
\begin{eqnarray}
\hat\Omega_{\hat q}(X)=
\left( \frac{2|\alpha |^2}{\pi}\right)^{1/4} 
\exp\left[-|\alpha |^2 (X -\underline{\hat q})^2 
+i|\alpha |^2 (\hat q^2 X+\hat q -2X)\right] \;.
\label{e:miniomegamir}
\end{eqnarray}                                               
The feedback shifts the object 
position by a quantity proportional to the output $X$. The operators 
$\hat\Omega_{\hat q}(X)$ becomes 
\begin{eqnarray}
\hat\Omega(X) =
\exp[i(\tau'/\tau)X\underline{\hat p}]\,\hat\Omega_{\hat q}(X)
\;,\label{e:omegafeed}
\end{eqnarray}      
where $\underline{\hat p}=\hat p/p_{\tau}$, with $p_{\tau}\doteq
\hbar l_{\tau}^{-1}$.
As the feedback is represented by a unitary operator, it does not modify
the POM, but changes only the state reduction, so that the posterior deviation 
$\sigma[\hat\varrho]^2 $ changes and condition (\ref{e:ONC}) 
can be beaten. 
\par Now we check if there is violation of the SQL. From Eq. 
(\ref{e:pomirsmall}), the precision of the measurement can be calculated 
as $\epsilon [\hat\varrho]^2 = l_{\tau}^2/(4|\alpha|^2)$, 
which is independent on the state of the system:
however, the position fluctuation depends on it. Thus, we must choose 
the object initial wave-function $|\psi\rangle $ at time $t=-\tau$
before the interaction, then calculate the reduced state after the first 
measurement at $t=0$, and then evolve it for a time interval $t_f$. 
We choose $\psi(q,-\tau)$ to be a MUW as in Eq. (\ref{e:muwinit}). 
The calculation can be made in two main steps. We first
multiply the function $\Omega_{\hat q}(X)$ by $\psi(q,-\tau)$              
(the reduction operator 
$\hat\Omega_{\hat q}(X)$ depends only on the position operator, and 
$\psi(q,-\tau)$ is written in the position representation). The second 
step consists in applying the feedback operator 
$\exp[i(\tau'/\tau)X\underline{\hat{p}}]$ to the result, 
shifting the $q$-coordinate 
by $l_{\tau}(\tau'/\tau)X$.
The reduced wave-function $\psi(q,0)$ turns out to be of the form
\begin{eqnarray}
&&\!\!\!\psi(q,0)\propto \exp\left\{ -q^2
\left[ {|\alpha |^2\over l_{\tau } ^2}\left( 1-iX\right) +
{1\over 4\delta _0^2}\right]\right.\\
&&\!\!\!\left. +2q\left[ {|\alpha |^2\over l_{\tau } }
\left( X+{i\over 2}\right) +
{q_0\over 4\delta _0^2}+{ik_0\over 2}-{|\alpha |^3\over l_{\tau } }
{\tau '\over \tau}X(1-iX)-{|\alpha |\over 4}{\tau '\over \tau}
{l_{\tau } X\over \delta _0^2}\right]  \right\}
\;.\nonumber\label{e:60}
\end{eqnarray}      
By remembering that the probe is prepared in a highly excited coherent 
state, and by keeping only the higher order terms in $|\alpha |$, 
we get
\begin{eqnarray}
\psi(q,0)\propto \exp\left\{-
{|\alpha |^2\over l_{\tau } ^2}\left( 1-iX\right)
\left(q+{\tau '\over \tau }|\alpha |l_{\tau } X\right)^2\right\}
\;.\label{e:61}
\end{eqnarray}      
The comparison between Eq. (\ref{e:61}) and Eqs. (\ref{e:tcs}) and 
(\ref{e:deltcs}) shows that $\psi(q,0)$ can be a TCS or not, 
depending on the 
value of the output $X$ of the measurement. In fact, 
the variance $\langle \psi |\Delta \hat q
^2(0)|\psi\rangle =l_{\tau } ^2/(4|\alpha |^2)$ can be set equal to 
$\hbar |\mu -\nu|^2/(2m\omega)$, with $|\mu|^2-|\nu|^2=1$, if 
$|\alpha |$ and $l_{\tau }$ are appropriately chosen. The time 
evolution for $\langle \psi |\Delta \hat q^2(t)|\psi\rangle $
is described by Eq. (\ref{e:deltcs}) with 
$\xi m\omega/(\hbar |\mu-\nu|^2)=-|\alpha |^2X/l_{\tau } ^2$, 
which means 
that $\xi \equiv \mbox{Im}(\mu^*\nu)=-X/2$. 
This implies that the reduced state (\ref{e:61}) is a TCS 
only when the measurement result gives a negative value $X<0$. 
However, this occurs only with $50\%$ probability, because 
the probability density for $X$ is Gaussian and centered in 
$X=0$. Hence, on the average---i. e. in Eq. (\ref{e:usql})---the SQL 
is not beaten. 
\par Even a pre-squeezing of the free mass before the measurement does 
not solve the matter in the hand. 
In fact, the operator $\hat\Omega(X)$ in (\ref{e:omegafeed}) would become
\begin{eqnarray}
\hat\Omega(X) =
\exp[i(\tau'/\tau)X\underline{\hat p}]\,\hat\Omega_{\hat q}(X)\,
\exp[-i(\tau''/\tau)(\underline{\hat q}\underline{\hat p}+
\underline{\hat p}\underline{\hat q})]
\;.\label{e:omegafeedsq}
\end{eqnarray}      
The squeezing operator 
$\exp[-i(\tau''/\tau)
(\underline{\hat q}\underline{\hat p}+\underline{\hat p}\underline{\hat q})]$ 
squeezes the $\hat q$-coordinate as 
$\hat q\to\hat q e^{(\tau''/\tau)}$. This reflects on the POM 
$d\hat\Pi_{\hat q}(X)$, which  changes to $d\hat\Pi_{\hat q 
e^{\tau''/\tau}}(X)$ so that the precision becomes 
$\epsilon^2 [\hat\varrho]=l_{\tau}^2/(4|\alpha|^2 e^{2\tau''/\tau})$. 
This precision becomes very small both if the initial state for the 
probe is very excited and if the pre-squeezing on the system is very 
high. However, the squeezing operator does not modify the functional 
form of the initial MUW, because it only rescales its average values: 
in fact $\exp[-i(\tau''/\tau)(\underline{\hat q}\underline{\hat
p}+\underline{\hat p}\underline{\hat q})]
|q\rangle =e^{\tau ''/2\tau }|e^{\tau ''/\tau }q\rangle $, 
which implies that $\psi(q,-\tau )\to \psi (q e^{\tau ''/\tau },-\tau)$. 
Thus, the reduced state of the system after the pre-squeezing, the 
measurement and the feedback is again of the form (\ref{e:61}), and 
the SQL is not overcome. 
This result suggests some considerations about the way of beating the SQL.
We know that, for generalized von Neumann measurements, the averaged 
precision equals the posterior deviation, which, if the condition
(\ref{e:pomunbias2}) is satisfied, coincides with the variance of the 
reduced wave-function averaged on the previous readout.
Thus, when starting from von Neumann schemes, in order to beat the SQL,  
we must change $\langle \Delta\hat q^2(0)\rangle $. 
We accomplished that by applying 
a pre-squeezing and a feedback to the system, and  we have seen 
that, if the probe is initially in a coherent state, at best the SQL can
be reached. The reason is that, even though the precision tends to zero 
for very high squeezing, the uncertainty of the reduced state does not 
decrease with time, in general. This suggests us to squeeze also the probe 
(\cite{oz88,walls93}) as we have
done in Section (\ref{s:gl}). This should reflect on the state of the 
free mass by reducing it to a contractive state, independently on the 
output of the measurement. Thus, we now calculate the 
operator $\hat\Omega$ when the probe is prepared in a squeezed coherent 
state $|-i|\alpha|,-r\rangle\equiv D(-i|\alpha |)
S(-r)|0\rangle $. The reason why we do not take a squeezed 
vacuum as in Ref. \cite{walls93} is apparent from the form of the 
evolution operator (\ref{e:evomir}): it represents a rotation 
in the complex phase space, and 
if applied to the vacuum it does not modify the state. The evaluation 
of the operator $\hat\Omega_{\hat q}(X)$ needs a quite long
derivation: we only report the result
\begin{eqnarray}
&&\!\!\!\!\hat\Omega_{\underline{\hat q}}(X)=
\left(\frac{2|\alpha |^2}{\pi \hat k^2}\right)^{\frac14} 
e^{-{i\over 2}(\hat{\phi}+\underline{\hat q})}\times\\
&&\!\!\!\!\!\!\!\!
\exp\left\{- i|\alpha |^2 \cos{\underline{\hat q}}
(2X -\sin{\underline{\hat q}})
-|\alpha |^2 (X -\sin{\underline{\hat q}})^2
\left[ {i\over \tan \underline{\hat q}}
\left( {1\over \hat k^2 e^{2r}}-1\right)+{1\over \hat k^2}\right]
\right\}\!\!\,,\nonumber\label{e:omegacs1}
\end{eqnarray}            
where
\begin{eqnarray}
\hat k^2&=&e^{2r}\sin ^2{\underline{\hat q}}+e^{-2r}
\cos ^2{\underline{\hat q}}\;,\\
\hat{\phi } &=&-\mbox{arctan}\left(e^{2r}\tan{\underline{\hat q}}\right)\;.
\end{eqnarray}
If we suppose that the mirror displacements are very small, which means 
that $\langle \hat q\rangle\ll l_{\tau}$, 
we have the first-order approximation for $\hat\Omega_{\underline
{\hat q}}(X)$
\begin{eqnarray}
&&\hat\Omega_{\underline{\hat q}}(X)=
\left(\frac{2|\alpha |^2 e^{2r}}{\pi }\right)^{\frac14}\\ 
&&\times\exp\left[i|\alpha |^2
\left(X\underline{\hat q}^2+\underline{\hat q}- 2X\right)
-|\alpha |^2 e^{2r}\left(X -\underline{\hat q}\right)^2 
+\frac{i\underline{\hat q}}{2}(e^{2r}-1)\right]\;.\nonumber
\label{e:omegacs2}
\end{eqnarray}      
As we have seen in the previous derivation, 
the pre-squeezing does not change the form of the
$\hat q$-representation of the system wave-function, but only 
rescales both the average and the variance by
squeezing factors. Hence, we choose an interaction described only 
by a measurement followed by a feedback, like in Eq. 
(\ref{e:omegafeed}). Once again, we first multiply the initial 
wave-function by $\hat\Omega_{\underline{\hat q}}(X)$, then shift 
$q$ by $|\alpha |l_{\tau}X/\tau $ and finally impose that $|\alpha |$
is very high. We get 
\begin{eqnarray}
\psi(q,0)\propto \exp\left\{-
{|\alpha |^2 e^{2r}\over l_{\tau } ^2}\left( 1-{iX\over e^{2r}}\right)
\left(q+{\tau '\over \tau} |\alpha |l_{\tau } X\right)^2\right\}
\;,\label{e:67}
\end{eqnarray}      
which is of the same functional form of the state reduction 
(\ref{e:61}), and is contractive only for negative values of the output 
$X$ of the measurement, that still occurs with $50\%$ probability. 
Therefore, by replacing the standard von Neumann Hamiltonian $\hat H_I=
\hat q\hat P$ with the interaction of the free mass with the 
radiation pressure, we can beat the SQL for certain values 
of the output, but on the average we only get a very narrow final state. 
This holds true even for a probe prepared in a
highly squeezed state, and using a pre-squeezing and a feedback.

\section{Conclusions}\label{s:con}

In this paper, we have engineered {\em ab initio} a measurement scheme
that allows to beat the SQL. We have shown that our scheme belongs to
a class of measurement models previously studied by Ozawa. 
The measurement can be performed in three-steps, involving a pre-squeezing
~stage, a von Neumann interaction $\hat H_{I}=\hat q\hat P$, and
a feedback. For large squeezing, and with the probe
prepared in a TCS, the state reduction puts the system into a TCS too,
and the SQL is breached. When the standard von Neumann interaction is replaced 
by a mirror-radiation interaction $\hat H_I=\hat q\hat A^{\dag}\hat A$, 
the SQL cannot be beaten, even though all other conditions
are kept the same. This is due to the fact that in the limit of small
free mass displacements the radiation-pressure Hamiltonian gives the
same probability distribution than the von Neumann one, but not the 
same state reduction. In fact, the von Neumann interaction has the
capability of transferring the shape of the wave-function from the probe
to the system, so that the reduced system state can be made contractive,
and thus the SQL is breached. On the other hand, this is no longer true
for the radiation-pressure Hamiltonian, where the state reduction of a
MUW is a very narrow state, which can be contractive with $50 \%$ 
probability, but on the average the SQL is not beaten. 
Thus we conclude that the experimental realization of the precise 
form of the von Neumann Hamiltonian is in order, if the measurement 
apparatus is designed to beat the SQL.

\end{document}